\documentclass[aps,superscriptaddress,groupedaddress]{revtex4}  
\usepackage{graphicx}  
\usepackage{dcolumn}   
\usepackage{bm}        
\usepackage{amssymb}   

\begin{document}


\title{The impact of nuclear masses near $N=82$ on $r$-process abundances}

\author{M. R. Mumpower}
\email{matt.mumpower@nd.edu}
\homepage{http://www.matthewmumpower.com}
\affiliation{Department of Physics, University of Notre Dame, Notre Dame, Indiana 46556, USA}
\affiliation{Joint Institute for Nuclear Astrophysics, USA}

\author{D. -L. Fang}
\affiliation{National Superconducting Cyclotron Laboratory, Michigan State University, East Lansing, MI 48824, USA}
\affiliation{Joint Institute for Nuclear Astrophysics, USA}

\author{R. Surman}
\affiliation{Department of Physics, Union College, Schenectady, NY 12308, USA}
\affiliation{Joint Institute for Nuclear Astrophysics, USA}

\author{M. Beard}
\affiliation{Department of Physics, University of Notre Dame, Notre Dame, Indiana 46556, USA}
\affiliation{Joint Institute for Nuclear Astrophysics, USA}

\author{A. Aprahamian}
\affiliation{Department of Physics, University of Notre Dame, Notre Dame, Indiana 46556, USA}
\affiliation{Joint Institute for Nuclear Astrophysics, USA}

\date{\today}

\begin{abstract}
We have performed for the first time a complete $r$-process mass sensitivity study in the $N=82$ region. 
We take into account how an uncertainty in a single nuclear mass propagates to influence important quantities of neighboring nuclei, including Q-values and reaction rates. 
We demonstrate that nuclear mass uncertainties of $\pm0.5$ MeV in the $N=82$ region result in up to an order of magnitude local change in $r$-process abundance predictions.
We identify key nuclei in the study whose mass has a substantial impact on final $r$-process abundances and could be measured at future radioactive beam facilities. 
\end{abstract}

\pacs{}
\maketitle

The NRC report ``Connecting Quarks to the Cosmos'' identified eleven of the most challenging open questions for all of physics in the 21st century. 
One of these eleven questions includes the identification of the site(s) for the production of the heaviest elements found in nature \cite{NRC2002}. 
Most of the elements above Fe are thought to have been produced by either the slow capture ($s$-process) or rapid capture ($r$-process) of neutrons in astrophysical environments \cite{BBFH1957}. 
The $s$ process proceeds close to stability and astrophysical sites have been identified, while the $r$ process allows the production of nuclei much further from stability and potential sites remain unresolved \cite{Arnould2007}. 
The $r$ process is of particular interest since it produces nuclei for which no experimental measurements currently exist and it must operate under extreme astrophysical conditions.  
One of the strongest motivations for the construction of new radioactive ion beam facilities world-wide has been the access to the properties of neutron-rich nuclei \cite{Schatz2013}. 
Much of our knowledge of short-lived nuclei remains incomplete, leaving the extrapolation of nuclear properties to various theoretical models with widely ranging assumptions and predictions. 

Despite these difficulties, progress is being made on the longstanding problem of the $r$-process astrophysical site. 
Recent ground-based and Hubble Space Telescope observations offer tantalizing evidence of a kilonova associated with a short-duration gamma-ray burst \cite{Berger2013}. 
The kilonova is presumably powered by the radioactive decay of $r$-process nuclei produced in a compact object merger, the likely progenitor of the burst. 
Current state-of-the-art nucleosynthesis simulations of the neutron-rich ejecta of mergers show robust production of heavy elements which undergo fission recycling 
\cite{Goriely2011,Korobkin2012}. 
Some nucleosynthesis of the heaviest elements is likely to come from supernova as they naturally explain the observations of $r$-process elements in metal-poor stars \cite{Sneden2008}. 
The latest state-of-the-art supernova calculations with neturino-driven winds do not yield production of `main' $r$-process elements $A\gtrsim110$ \cite{Hudepohl2010,Fischer2010}. 
Instead, current calculations show a `weak' $r$-process, $A\lesssim110$, with variability in final abundances based off progenitor input \cite{Arcones2013}. 
This environment is incredibly complex to model, requiring a full inclusion of neutrino physics before a complete picture arises \cite{Roberts2012}.

Simulations of the $r$ process rely on calculations of individual nuclear physics properties far from stability which inherently contain large uncertainties and correlations. 
In an attempt to disentangle the specific nuclear properties that significantly impact the outcome of a given astrophysical scenario, studies of individual nuclear masses, binding energies, neutron separation energies, $\beta$-decay rates, and neutron capture rates have been performed in the context of both supernova and neutron star merger $r$-process scenarios \cite{Surman2009,Beun2009,Mumpower2012c,Brett2012,Suzuki2012,Nishimura2012,AIP-BE,AIP-Beta,AIP-NC}. 
All of these investigations conclude that the details of the final abundance pattern depend critically on the individual nuclear properties of nuclei along the $r$-process path and nuclei populated as material moves back to stability during last stage of the $r$ process known as freeze-out. 
For a main $r$ process these important nuclei are found to lie near the $N=82$ and $N=126$ closed shells and in the rare earth region $A\sim160$. 
The sensitivity of abundances to individual nuclear properties has been explored over a range of astrophysical conditions and nuclear models which shows the robustness of these conclusions \cite{INPC2013,ICFN5}.

Nuclear masses are particularly important for the $r$ process as they enter into the calculations of all of the aforementioned nuclear properties. 
Therefore, in order to obtain a more complete picture of the impact of individual masses on $r$-process abundances, a new approach is warranted which combines the techniques of previous studies. 
We recalculate Q-values, neutron capture rates and weak decay properties self-consistently when the uncertainty in a single nuclear mass is studied. 
The goal of this work is to show how uncertainties in an individual nuclear mass propagates to influence important quantities in 
neighboring nuclei and ultimately change final abundances in the context of a hot $r$-process wind. 
The results of this approach are multi-fold: (1) We identify the most important nuclear masses to measure, (2) we can put direct constraints on precision needed for measurements of nuclear properties and theoretical models, and (3) longterm, we can use the previous two points to constrain $r$-process models.

To investigate the impact of uncertain nuclear masses in the $N=82$ region on $r$-process abundances we use the concept of a sensitivity study. 
A sensitivity study begins with the choice of astrophysical conditions and nuclear inputs which is called the `baseline' simulation. 
We then vary one mass at a time, recalculate the nuclear properties that depend on the mass, rerun the $r$-process simulation, and compare it to the baseline. 
The change between the abundances of these two simulations measures the extent of influence the individual nuclear mass has on the predicted abundance pattern. 
This procedure complements other methods in the literature which quantify correlations between nuclear physics inputs in the context of astrophysical environments \cite{Bertolli2013}.

For our baseline simulation we choose a hot wind $r$-process. 
In a hot wind an equilibrium is established between neutron captures and their inverse reaction, photo-dissociation, and the $\beta$-decays that move the nuclear flow to higher atomic number, $Z$, control the timescale for heavy element production. 
Nuclear masses are particularly important during equilibrium as they directly set the $r$-process path, or set of most abundant isotopes, for a given 
temperature and density \cite{BBFH1957}. 
Once the supply of free neutrons is consumed the $r$-process path begins to move back to stability. 
The criterion of neutron exhaustion signals the start of the freeze-out phase of the $r$ process in which key abundance features are formed, such as the rare earth peak 
\cite{Mumpower2012b}. 
Additional neutrons during this time come from photo-dissociation, neutrons emitted promptly after $\beta$-decay, or by fission \cite{Mumpower2012a}. 

Hot $r$ process winds have been suggested to occur in a number of environments including supernova, neutron star mergers or black hole accretion disks. 
Here we employ a parameterized model from \cite{Meyer2002}. 
This trajectory has an entropy of $200$ $k_B$, an electron fraction of $Y_e=0.3$ and a timescale of $80$ ms yielding the production of heavy elements which extends beyond the region of interest but is not neutron rich enough to consider fission recycling. 
We use a dedicated $r$-process reaction network code \cite{Surman1997,Surman2001} updated for recent studies \cite{Beun2008,Surman2008,Mumpower2012c}.

To provide a consistent basis for our calculation of nuclear properties we use FRDM masses \cite{FRDM1995}. 
This model has an root-mean-square (rms) error around $0.65$ MeV compared to known masses from the latest Atomic Mass Evaluation (AME) \cite{AME2012}. 
We use these theoretical masses for all nuclei in our network since an artificial jump from theory to experiment may be introduced by attempting to supplement with measured masses. 
This is why our results show sensitivities for known masses in the AME. 

\begin{figure}
 \begin{center}
  \centerline{\includegraphics[width=140mm]{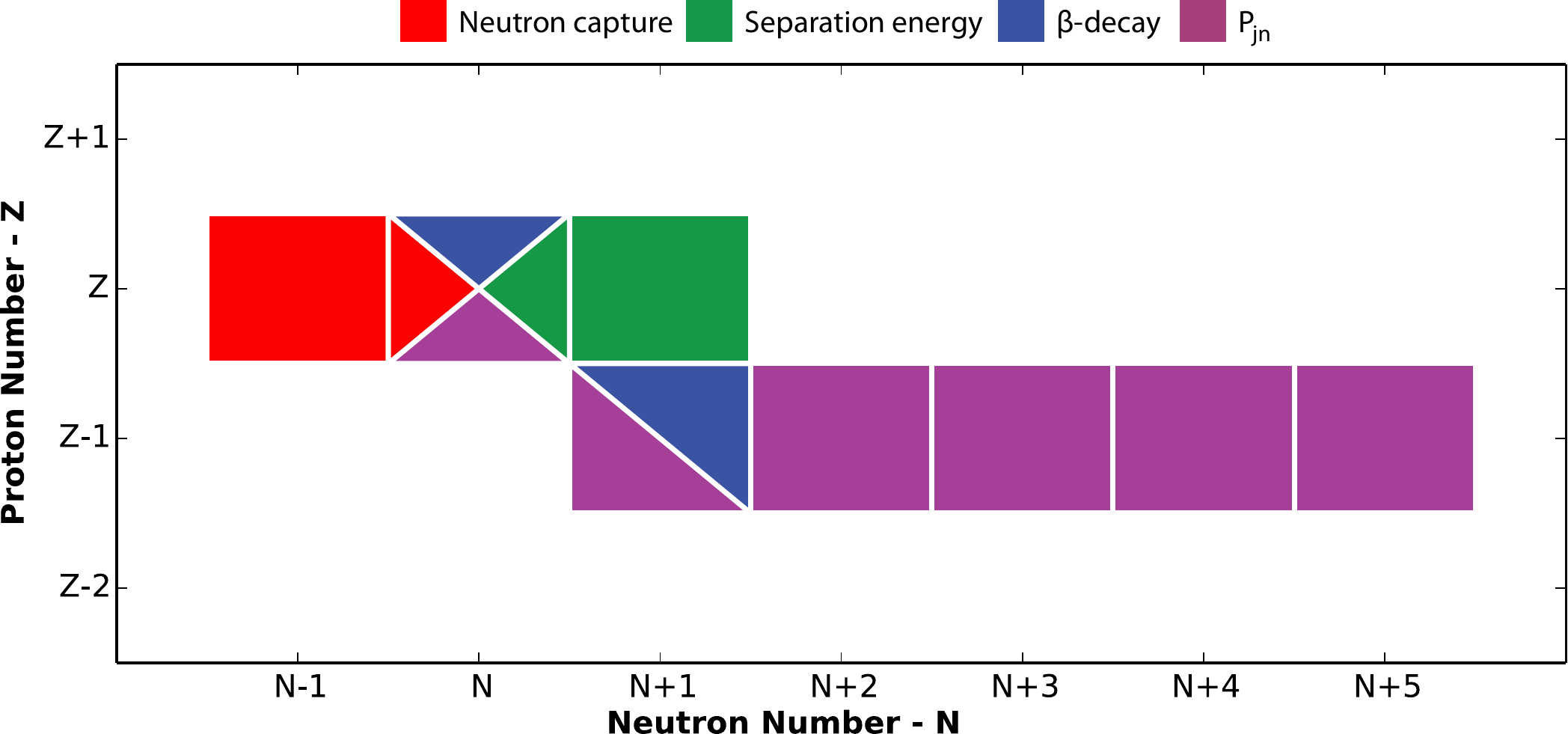}}
  \caption{\label{fig:mods} Shows the quantities of neighboring nuclei of importance to the $r$-process that are altered by a change in mass of nucleus with $Z$ protons and $N$ neutrons.}
 \end{center}
\end{figure}

When a nuclear mass is varied we recalculate all the relevant nuclear properties of neighboring nuclei that depend on the changed mass. 
Specifically, if the mass of a nucleus ($Z$,$N$) with $Z$ protons and $N$ neutrons is varied then it leads to changes in the neutron capture rates of ($Z$,$N$) and ($Z$,$N-1$), the separation energies of ($Z$,$N$) and ($Z$,$N+1$), the $\beta$-decay rates of ($Z$,$N$) and ($Z-1$,$N+1$), and $\beta$-delayed neutron emission probabilities of ($Z$,$N$), ($Z-1$,$N+1$), ($Z-1$,$N+2$), ($Z-1$,$N+3$), ($Z-1$,$N+4$) and ($Z-1$,$N+5$) as shown in Fig. \ref{fig:mods}.

Nuclear masses go into the calculation of neutron capture rates via the dependency on neutron separation energy and reduced mass. 
Depending on the model, the separation energy can enter into the optical potential, the gamma strength function, particle and $\gamma$ transmission coefficients and level density. 
Note that that a change to the separation energy of the target nucleus does not propagate to the capture rate calculations. 
For example, the capture rate of ($Z$,$N+1$) does not change when the mass of ($Z$,$N$) is varied since this mass variation does not alter the separation energy of the compound nucleus ($Z$,$N+2$). 
Across the chart of nuclides we use neutron capture rates calculated with the publicly available statistical model code, TALYS \cite{TALYS} with photo-dissociation rates calculated by detailed balance. 
To propagate the changes to the neighboring neutron capture rates we invoke the `massnucleus' command in TALYS.

The dependence of weak decay properties on nuclear masses can be derived from following formula \cite{Moller2003}:
\begin{eqnarray}\label{eqn:weak}
\lambda_\beta\equiv\frac{\ln(2)}{t_{1/2}}=\sum_{i} f^I_{\omega_i} C^{I}(\omega_i)
\end{eqnarray}
here $\lambda_\beta$ is the $\beta$-decay rate, $t_{1/2}$ is the half-life, $i$ denotes the $i^{\text{th}}$ excited state of the daughter nucleus with energy 
$E_i$, $\omega_i=(Q_\beta-E_i)/m_e$ is the $\beta$-decay energy to this state in units of electron mass, $I$ is the type of the decay, either Gamow-Teller 
(GT) or First-Forbidden (FF), $f$ is the phase space factor and $C$ is the $\beta$-strength function. 
Approximating the phase space factor, $f(\omega)\sim \omega^5$ for allowed $\beta$-decay, one can deduce the general dependence of the half-life on $\beta$-decay energies. 
For the most neutron-rich $r$-process nuclei, the $\beta$-decay Q-values are above $10$ MeV. 
This means uncertainties in nuclear masses on the order of the RMS error of FRDM could bring out an overall change of the Q-value to about $1$ MeV, which in turn could change the half-life by a factor of two or more. 
The dependence of the $\beta$-delayed $j$-neutron emission probabilities $P_{jn}$ on nuclear masses is less straight forward, as the change in mass also leads to a change of the neutron separation energies. 
To estimate the unknown $\beta$-decay rates and corresponding neutron emission probabilities of neutron-rich nuclei we use a combination of FRDM based calculations. 
Since the mass changes need to be propagated to the $\beta$-decay rates in the $N=82$ region we use newer QRPA solutions from \cite{DLF2013} and outside this region we use rates from \cite{Moller2003}. 
Variations in nuclear masses of the range considered here have a small effect on the matrix elements of QRPA calculations and so we only recalculate the phase space factor.

To gauge the impact of a mass uncertainty on the final abundances we compute the metric
\begin{equation}\label{eqn:F}
F=100\sum_{A}|X(A)-X^{b}(A)|
\end{equation}
where $X^{b}(A)$ is the final isobaric mass fraction in the baseline simulation, $X(A)$ is the final isobaric mass fraction of the simulation when all nuclear 
inputs have been modified based off the change in a single mass, and the summation runs over the entire baseline pattern. The isobaric mass fractions are defined by 
$X(A)\equiv\sum\limits_{Z+N=A}X(Z,N)$ where $X(Z,N)$ is the mass fraction of a species in our network given $Z$ protons and $N$ neutrons. We note that the 
total mass fraction always sums to unity and an isobaric mass fraction can be transformed into an abundance by the relation $X(A)=AY(A)$.

\begin{figure}
 \begin{center}
  \centerline{\includegraphics[width=140mm]{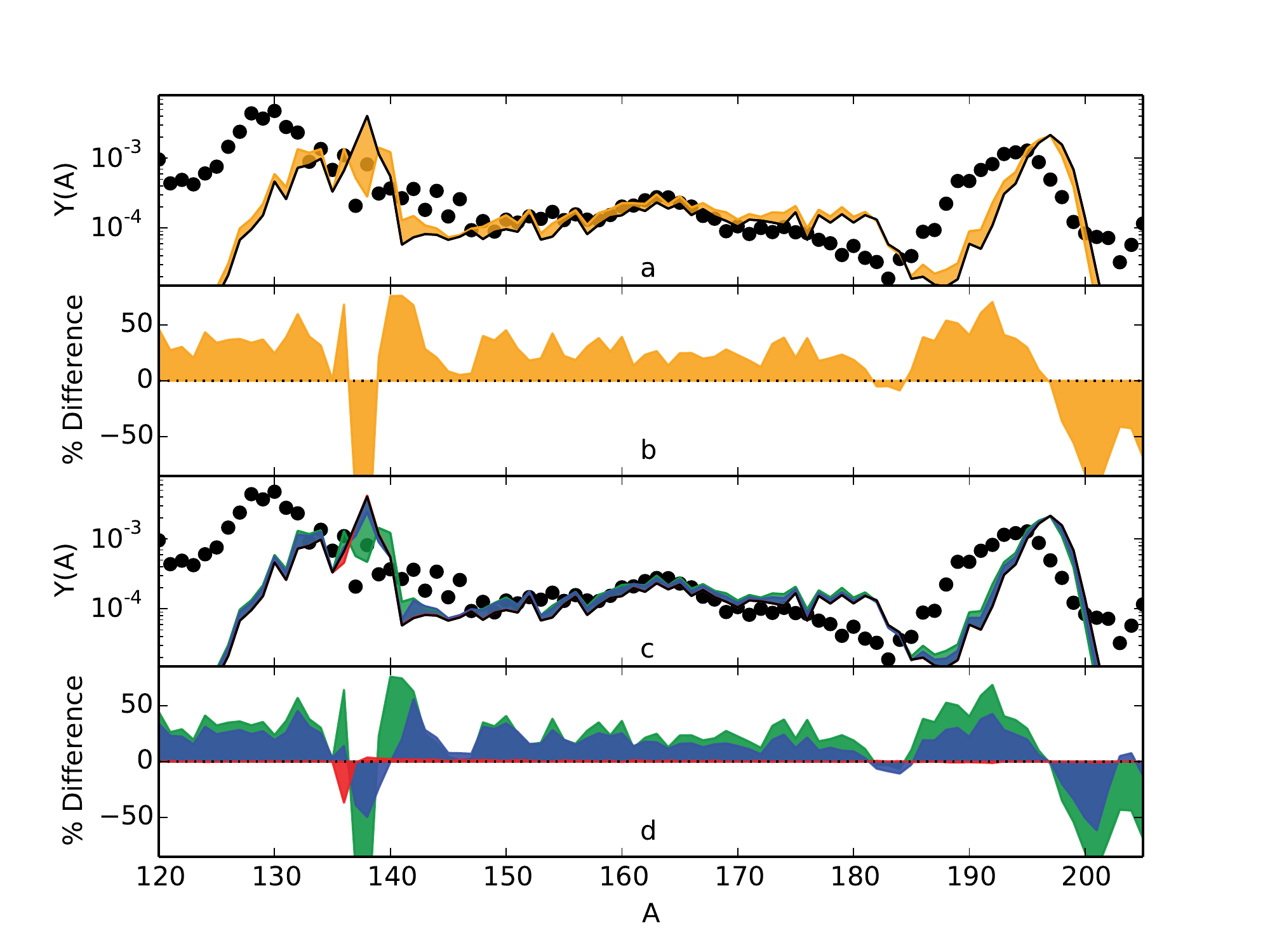}}
  \caption{\label{fig:ab} (a) Shows the change in final $r$-process abundances when the mass of $^{140}$Sn is increased by $0.5$ MeV and all dependencies have been considered compared to baseline simulation (black) and solar data (black circles) from \cite{Arlandini1999}. (b) Percent difference of the abundances in (a) to baseline. (c) Separate simulations show the contribution of the mass variation when only propagating changes to neutron captures (red), $\beta$-decays (blue) and separation energies (green) as in Fig. \ref{fig:mods}. (d) Percent difference of the abundances in (c) to baseline.}
 \end{center}
\end{figure}

We now show the effect of an individual mass variation on the final $r$-process abundance pattern for the case of $^{140}$Sn. In panel (a) and (b) of Fig. 
\ref{fig:ab}, we show the changes to the final abundance pattern of our baseline simulation that result from adding $0.5$ MeV to the mass of $^{140}$Sn 
($Z=50$) and propagating changes to all 12 neighboring nuclei affected by it as shown in Fig. \ref{fig:mods}. We additionally performed three auxiliary simulations to 
highlight the impact of this mass variation to each of the neighboring nuclear properties. We propagated changes to the neutron capture rates only, the 
$\beta$-decay rates and associated $P_{jn}$ values only, and the one-neutron separation energies only, with the results shown in panels (c) and (d) of Fig. 
\ref{fig:mods}. The latter, shown in green, is the same type of calculation used in Ref. \cite{Brett2012}, except here we use different $\beta$-decay rates and 
$r$-process conditions. In what follows we analyze the separate cases to gain insight into the physical mechanisms at play during the simulation where all of 
the properties change.

The change in final abundances seen in panel (a) is dominated by the shift in separation energies (green line in panel (c) of Fig. \ref{fig:ab}). 
In the baseline simulation, $^{140}$Sn is populated during equilibrium and lies along the $r$-process path for roughly 1 second. Thus, an increase to the mass of this 
nucleus results in the primary effect of shifting the equilibrium path further out to $^{142}$Sn. Since the $r$-process path is more neutron-rich it 
substantially speeds up the consumption of free neutrons and quickly moves back towards the lighter Tin isotopes.

A smaller change to the abundance pattern comes from $\beta$-decay, as can be seen from the blue line in panel (c) of Fig. \ref{fig:ab}. 
In the baseline simulation the flow out of $^{140}$Sn is controlled by the 1-neutron emission decay channel. 
A mass increase of $0.5$ MeV to $^{140}$Sn lowers the half-life of this nucleus by 45\% which results in a speed up of the 1-neutron emission decay channel and faster nuclear flow through the region. 
In this case, we find that changes to $P_{jn}$ values have only a minor impact on the nuclear flow. 

The change to neighboring neutron capture rates is the smallest effect, shown in the red line in panel (c) of Fig. \ref{fig:ab}, since neutron capture rates 
are only important out of equilibrium. The additional mass to $^{140}$Sn increases both the capture rates of $^{140}$Sn and and $^{139}$Sn in the temperature 
range under consideration, $T\lesssim2$GK. The result is that when only neutron capture rate changes are considered, a small neutron capture effect is seen 
which moves material to higher atomic mass at late times\cite{Mumpower2012c}. For some nuclei, e.g. $^{142}$Sn, under these conditions this is the dominant 
effect from a change in mass. We note that the uncertainty in neutron capture rates is underestimated by this approach. Large uncertainties in neutron capture 
rates are influenced by choice of gamma strength functions and level density models \cite{Beard2014}.

With an addition of $0.5$ MeV to the mass of $^{140}$Sn the value of $F$ is $31.24$ when all effects are propagated, one of the highest sensitivities of any 
nucleus in our study. When the variation in this mass is propagated individually it yields $F_{(n,\gamma)}=1.4$ for the neutron capture effects, 
$F_{\beta}=16.0$ for weak decay effects and $F_{(\gamma,n)}=29.96$ for equilibrium path and photo-dissociation effects. This means that in the simulation 
where all nuclear properties are updated, the shift in the equilibrium path to $^{140}$Sn from the change in separation energies of $^{140}$Sn and $^{141}$Sn 
dominates. Due to this shift $^{140}$Sn never lies along the $r$-process path, and so the remaining effects are reduced, similar to saturation discussed in 
Ref. \cite{Mumpower2012c}.

\begin{figure}
 \begin{center}
  \centerline{\includegraphics[width=160mm]{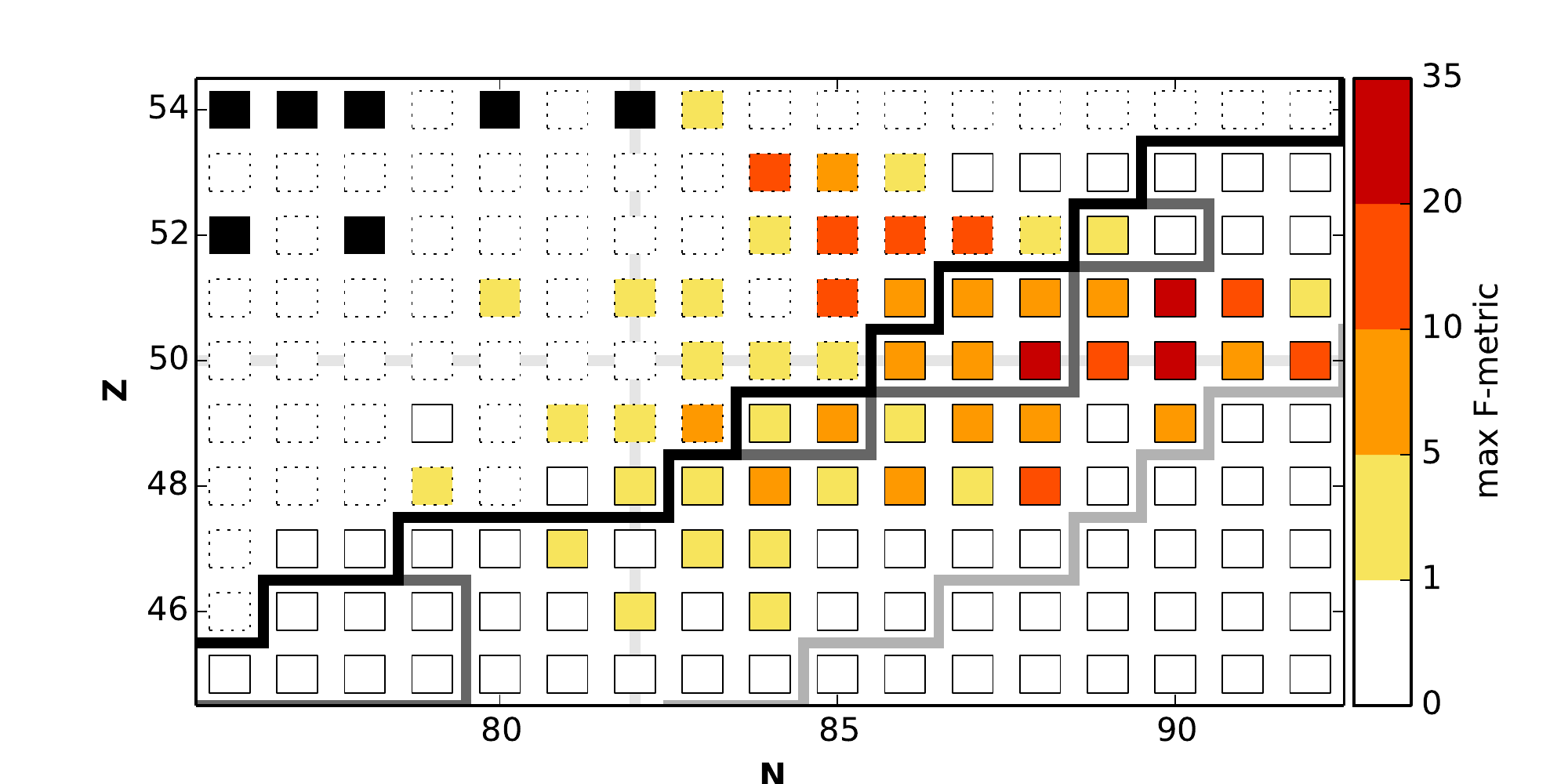}}
  \caption{\label{fig:fgrid} Nuclei from this study that significantly impact final $r$-process abundances in the $N=82$ region with an uncertainty of $\pm0.5$ MeV. The black line shows the extent of measured masses from the AME 2012. Nuclei with masses measured within $0.1$ MeV are denoted by a dotted border and those greater denoted by solid border. Accessibility limits shown for CARIBU (dark gray) and FRIB (light gray).}
 \end{center}
\end{figure}

The effects of individual variations in 170 nuclear masses on $r$-process abundances are summarized in Fig. \ref{fig:fgrid}. 
The calculations were carried out as outlined above for $^{140}$Sn. 
Here we have taken the maximum $F$-value between the case where $0.5$ MeV was added to or subtracted from the FRDM mass. 
Qualitatively, the distribution of important nuclei agrees with previous work \cite{Brett2012}. 
We find the most important nuclear masses are for nuclei along the equilibrium $r$-process path as well as along the decay pathways back to stability, most of which lie within experimental reach of CARIBU and FRIB. 
In this work we find that sensitivities of $r$-process abundances to nuclear masses increases when all properties are propagated consistently due mainly to the dependence of weak decay properties on nuclear masses. 
Our distribution of important nuclei favors the region to the right of the $N=82$ closed shell due the position of the $r$-process path and the amount of late-time neutron capture occurring in this environment \cite{Arcones2011,Mumpower2012b}. 
More sensitivity would be seen for nuclei close to the $N=82$ closed shell if we considered different astrophysical conditions. In the case of a cold wind or 
neutron star merger $r$-process, equilibrium and photo-dissociation effects would play a less prominent role for most nuclei. This is because the 
photo-dissociation rates are likely frozen out by the time main $r$-process nuclei are populated and so the changes to separation energy would only affect 
neutron captures and weak decay properties. In both of these environments we expect the $\beta$-decay contribution to be similar in size to the neutron capture 
component owing to the fact that the $r$-process path during freeze-out is supported by a balance between $\beta$-decay and neutron capture 
\cite{Wanajo2007,Mumpower2012a}.

In summary, we have shown how uncertainties in \textit{individual} nuclear masses propagate to influence and shape the $r$-process abundance distribution near the $N=82$ closed shell. 
To propagate uncertainties in a single nuclear mass we recalculate Q-values, neutron capture rates, photo-dissociation rates, $\beta$-decay rates and $\beta$-delayed neutron emission probabilities, as shown in Fig. \ref{fig:mods}. 
We have shown that mass uncertainties of $\pm0.5$ MeV can produce large local changes, on the order of 100\% difference, and smaller global changes, on the order of 20-40\% difference, to $r$-process abundances. 
In terms of our metric, a value of $F\sim30$ for a mass variation in a single nucleus can yield over an order of magnitude local change in the $r$-process abundance predictions. 
In the context of a hot wind $r$-process we have shown that shifts in the equilibrium path and changes to the to weak decay properties play an essential role. 

Furthermore, it had been proposed in Ref. \cite{Bohigas2002} that mass models have some inherent chaotic noise on the order of $0.2$ MeV, however Ref. \cite{Barea2005} showed that in fact no such chaotic limit exists and higher precisions well below $0.1$ MeV are achievable in the calculation of nuclear masses if all the physical ingredients of nuclei are incorporated into the model. 
We have explored mass uncertainties as low as $\pm0.1$ MeV, which are smaller than the rms values of all leading mass models. 
In this case, we find the same nuclei from this study still impact the final $r$-process abundances except with reduced F-values that reach a maximum of $F\sim15$ for the most influential nuclei. 
Thus, to capture the finer details of solar $r$-process abundance signatures it is important to strengthen efforts in the direction of reducing uncertainties in mass models. 
We plan to extend these calculations beyond the $N=82$ region and consider a range of astrophysical environments and nuclear models in a forthcoming publication. 

We thank Gail McLaughlin for helpful discussions. 
This work was supported by the NSF through the Joint Institute for Nuclear Astrophysics grant numbers PHY0822648 and PHY1419765.

\bibliographystyle{unsrt}
\bibliography{refs}

\end{document}